\pdfoutput=1
\documentclass{article}
\usepackage{spconf,amsmath,graphicx}
\usepackage{todonotes}
\usepackage{caption}
\usepackage{multirow}
\usepackage{subcaption}
\usepackage{makecell}
\usepackage{tabu}
\graphicspath{ {./Figures/} }

\title{INTERPRETING CNN FOR LOW COMPLEXITY LEARNED SUB-PIXEL MOTION COMPENSATION IN VIDEO CODING}

\name{Luka Murn$^{\star}$$^{\dagger}$, Saverio Blasi$^{\star}$, Alan F. Smeaton$^{\dagger}$, Noel E. O'Connor$^{\dagger}$, Marta Mrak$^{\star}$\thanks{The work described in this paper has been conducted within the project JOLT funded by the European Union’s Horizon 2020 research and innovation programme under the Marie Skłodowska-Curie grant agreement No 765140}}

\address{
  $^{\star}$BBC Research and Development, London, UK \\
  $^{\dagger}$Dublin City University, Dublin, Ireland}

\begin{document}
\ninept
\maketitle
\begin{abstract}
Deep learning has shown great potential in image and video compression tasks. However, it brings bit savings at the cost of significant increases in coding complexity, which limits its potential for implementation within practical applications. In this paper, a novel neural network-based tool is presented which improves the interpolation of reference samples needed for fractional precision motion compensation. Contrary to previous efforts, the proposed approach focuses on complexity reduction achieved by interpreting the interpolation filters learned by the networks. When the approach is implemented in the Versatile Video Coding (VVC) test model, up to $4.5\%$ BD-rate saving for individual sequences is achieved compared with the baseline VVC, while the complexity of learned interpolation is significantly reduced compared to the application of full neural network.
\end{abstract}
\section{Introduction}
\label{sec:intro}
Advanced video compression solutions, such as the current state-of-the-art High Efficiency Video Coding (HEVC) standard \cite{hevc} and the next-generation Versatile Video Coding (VVC) \cite{bross2019vvcdraft} standard, rely on the investigation of new, more efficient compression tools. In order to even further reduce the bitrates necessary to transmit content at higher video qualities, solutions based on learned methods, rather than traditional, hand-crafted video coding methods are being explored. In this context, deep learning schemes similar to ones proven to be useful in image processing tasks, are showing great potential in video coding applications as well. Methods based on Convolutional Neural Networks (CNNs) provide significant improvements in tasks such as image denoising \cite{denoise}, image super-resolution \cite{srcnn} and image colourisation \cite{Blanch_2019}. For these reasons, significant research efforts have been focused on ways to integrate CNN-based solutions into next generation video coding schemes \cite{DongReview,SiweiReview, Santamaria_2018}. 

When used in video coding for higher compression, such solutions have shown to bring coding gains at the cost of significant increases in complexity and memory consumption. In many cases, the high complexity of these schemes, especially on the decoder side, limits their potential for implementation within practical applications. Nevertheless, schemes based on highly simplified neural network (NN) models have been proposed \cite{Westland_2019}, while some have been adopted into the latest VVC drafts, including Matrix Intra-Prediction (MIP) modes \cite{intradeep} and Low-Frequency Non Separable Transform (LFNST) \cite{lfnst2019koo,zhao2016nsst}. 

Most modern video coding solutions rely on sub-pixel (fractional) Motion Compensation (MC) to refine integer motion vectors and provide more accurate prediction samples. The reference samples are interpolated by means of fixed N-tap filters which are sequentially applied in the horizontal and vertical direction to produce fractional samples. VVC inherits the same 8-tap filter to generate half-pixel samples and 7-tap filters for quarter-pixel samples \cite{KemalFilters} as in HEVC, but extends these with filters that provide up to sixteenth-pixel precision samples as well as an alternate half-pixel filter. However, these fixed filters may not describe the original content well enough or capture the diversity within the video data.  

In this paper, a novel tool based on NNs is presented that improves the interpolation of reference samples needed for fractional precision MC. Contrary to previous NN-based efforts, the proposed approach focuses on complexity reduction which is achieved by interpreting the results learned by the networks. In his context, interpretability aims to understand the relationships learned by a NN, facilitating the derivation of simple algorithms from a multi-layer network. Fractional interpolation models obtained this way preserve the advantages of the learned models, while enabling their low-complexity implementation.

\section{State of the art}
\label{sec:state}
An approach to using super-resolution CNNs to generate half-pixel interpolated fractional samples was introduced in \cite{Yan2017}, reporting $0.9\%$ Bjøntegaard delta-rate (BD-rate) \cite{bjontegaard2001calculation} reductions under low-delay P (LDP) configuration when replacing HEVC luma filters. Training separate networks for luma and chroma channels was presented in \cite{ChromaMC}. The resulting models were integrated within the HEVC reference software as a switchable interpolation filter, achieving $2.9\%$ BD-rate coding gains under the LDP configuration.

As a follow-up to \cite{Yan2017}, Yan {\em et al}. proposed to formulate sub-pixel MC as an inter-picture regression problem rather than an interpolation problem \cite{Yan2019}. The resulting method uses $15$ networks, one for each quarter-pixel fractional shift. The input to each network was the decoded reference block for that position, where the ground truth was the original content of the current block. Different NNs were trained for uni-prediction and bi-prediction and for different QP ranges, resulting in a total of $120$ NN-based interpolation filters. Two NN structures were compared when training the NNs, a $3$-layer structure referred to as Super-Resolution CNN (SRCNN), and a deeper model with multiple branches based on Variable-filter-size Residue learning CNN (VRCNN), as proposed in \cite{vrcnn}. When tested on $32$ frames, $2.9\%$ BD-rate gains were reported for VRCNN under LDP configuration with respect to HEVC, with $2.2\%$ for SRCNN.

While these methods consistently improve the efficiency of video compression by providing more accurate sub-pixel interpolated samples, they have high complexity requirements to produce CNN-based estimations. The SRCNN model implemented as a switchable interpolation filter resulted in an almost $50$ times higher decoder run-time compared to the HEVC anchor, while VRCNN increased the run-time by more than $200$ times \cite{Yan2019}. New solutions to reduce the complexity of these models would be highly beneficial to ensure such methods can be integrated within practical coding solutions.

Interpreting and understanding relationships learned by the network enables the derivation of streamlined, less complex algorithms which achieve similar performance to the original models. In \cite{Murdoch_2019}, a framework for defining machine learning interpretability methods was introduced. Interpretability could be achieved using model-based methods prior to training, by defining a network structure that is simple enough to be analytically understood, while sophisticated enough to fit underlying data. Interpretability can also be achieved using post-hoc methods, by analysing the NN models after training, providing valuable insights into the learned relationships between inputs and outputs. 

The approach proposed in this paper builds on the algorithms in \cite{Yan2019}, with the goal of reducing the complexity of SRCNN-based sub-pixel MC using interpretability of learned NN models. Both model-based and post-hoc interpretability methodologies are employed with the goal of capturing how individual features of the input data contribute to the output predictions, thus deriving simple yet accurate predictions.

\section{Proposed approach}
\label{sec:approach}

The SRCNN model presented in \cite{Yan2019} contains $64$ individual $9\times9$ convolutional kernels in the first layer, $32$ individual $1\times1$ kernels in the second layer, and $32$ individual $5\times5$ kernels in the final layer. It is worth mentioning that the output of the network $\mathbf{Y}$ (motion copensated prediction) is modified by adding the input $\mathbf{X}$ (reference samples), which means the output of the final convolutional layer is formed of prediction residuals $\mathbf{R}$. In the machine learning context, residuals are defined as the difference between the output and the input, formally $\mathbf{Y} = \mathbf{R} + \mathbf{X}$. 

\begin{figure}
    \centering
    \includegraphics[width=0.4\textwidth]{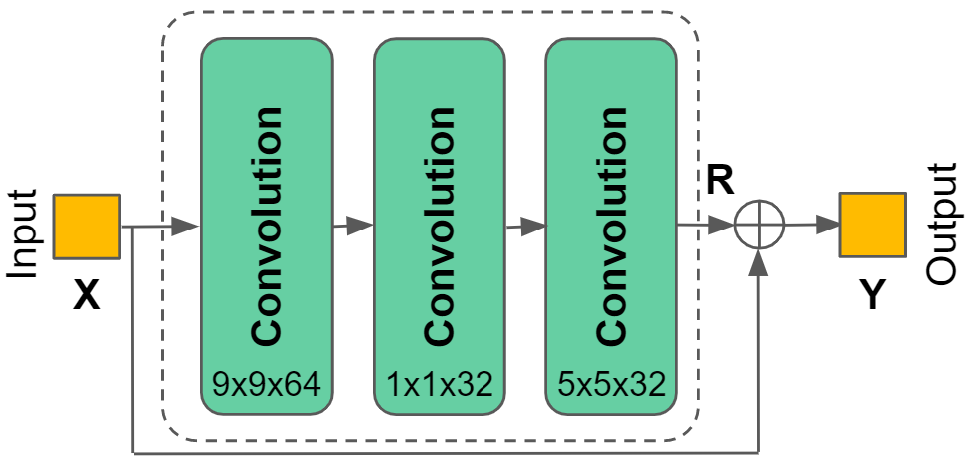}
    \caption{$ScratchCNN$ network architecture}
    \label{fig:scratchcnn}
\end{figure}

Following a model-based interpretability approach, a new simplified structure can be defined by removing activation functions and biases from the network, as they introduce non-linearities between layers which do not allow simplifications. The original $3$-layer SRCNN network contains ReLU activation functions after the first and second layer, while biases are added to weighted inputs of each layer. The removal of non-linearities does not affect the network performance, as discussed in Section \ref{sec:results}. The proposed SRCNN without ReLUs and biases, referred to as ScratchCNN, is illustrated in Fig.~\ref{fig:scratchcnn}.

\begin{figure*}
    \centering
    \includegraphics[width=\textwidth]{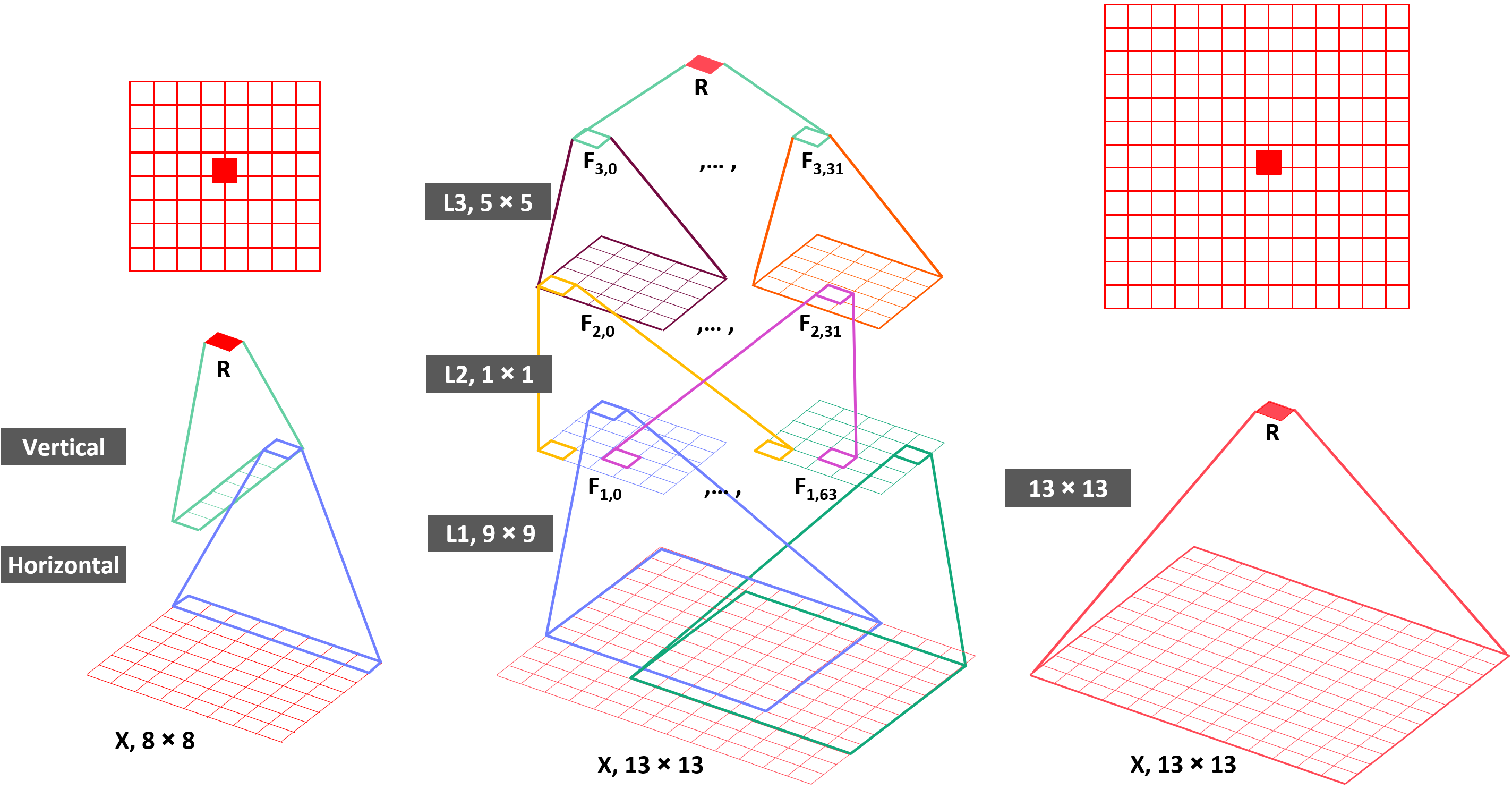}
    \caption{Fractional pixel derivation process for VVC (left), NN interpolation filter (centre) and proposed approach (right). VVC requires $8\times8$ samples (top-left) to predict a pixel; NN and proposed approach require $13\times13$ samples (top-right).}
    \label{fig:interpret}
\end{figure*}

The ScratchCNN training process is outlined in Section~\ref{sec:data-training}. Once a trained model is available, post-hoc interpretability can be applied to derive a simple interpolation filter. As seen in Fig.~\ref{fig:interpret}, the first convolutional layer output, $\mathbf{F}_1$, is obtained from a given input $\mathbf{X}$ as:

\begin{equation}
\label{eq:firstlayer}
\mathbf{F}_{1,i} = \mathbf{K}_{1,i} * \mathbf{X},
\end{equation}

\noindent where $\mathbf{K}_{1,i}$ correspond to $9\times9$ convolutional kernels and $i=0,..,63$. Second convolutional layer output, $\mathbf{F}_2$, is obtained as:

\begin{equation}
\label{eq:secondlayer}
\mathbf{F}_{2,j} = \displaystyle\sum_{i=0}^{63} K_{2,j} \mathbf{F}_{1,i},
\end{equation}

\noindent where $K_{2,j}$ correspond to $1\times1$
convolutional kernels, i.e. scalar values, and $j=0,..,31$. The final convolutional layer output $\mathbf{R}$ is obtained from $\mathbf{F}_3$ feature maps as:

\begin{equation}
\label{eq:thirdlayer}
\mathbf{F}_{3,j} = \mathbf{K}_{3,j} * \mathbf{F}_{2,j}, 
\end{equation}

\noindent and their summation for each $j=0,..,31$ as:

\begin{equation}
\label{eq:output}
\mathbf{R} = \displaystyle\sum_{j=0}^{31} \mathbf{F}_{3,i}.
\end{equation}

Additionally, unlike the networks described in \cite{Yan2019} which apply zero padding between layers to keep the input size consistent, none of the convolutional layers in the proposed simplified CNN apply padding. They instead use available samples from the reference frame. Thus, the input $\mathbf{X}$ is extracted into patches only prior to the first convolutional layer. As a $5\times5$ convolution is applied on top of a $9\times9$ convolution, then $13\times13$ reference samples have to be considered. Values at input positions $X_{i,j}$, where $i,j=0,..12$, are multiplied with several convolutional kernel weights per layer. Summing all the weights with which a $X_{i,j}$ has been multiplied with, leads to a $13\times13$ matrix $\mathbf{M}$ created from trained CNN, described as:

\begin{equation}
\label{eq:interpret}
\mathbf{R} = \mathbf{M} * \mathbf{X} = \mathcal{F}(\mathbf{K}_3, K_2, \mathbf{K}_1, \mathbf{X}).
\end{equation}

A non-separable $2$D filter $\mathbf{M}$ is obtained. The filter coefficients represent the contribution of each of the reference samples in a fixed $13\times13$ window surrounding the interpolated fractional sample, as shown on the top-right of Fig.~\ref{fig:interpret}.

\begin{figure}
    \centering
    \includegraphics[width=0.4\textwidth]{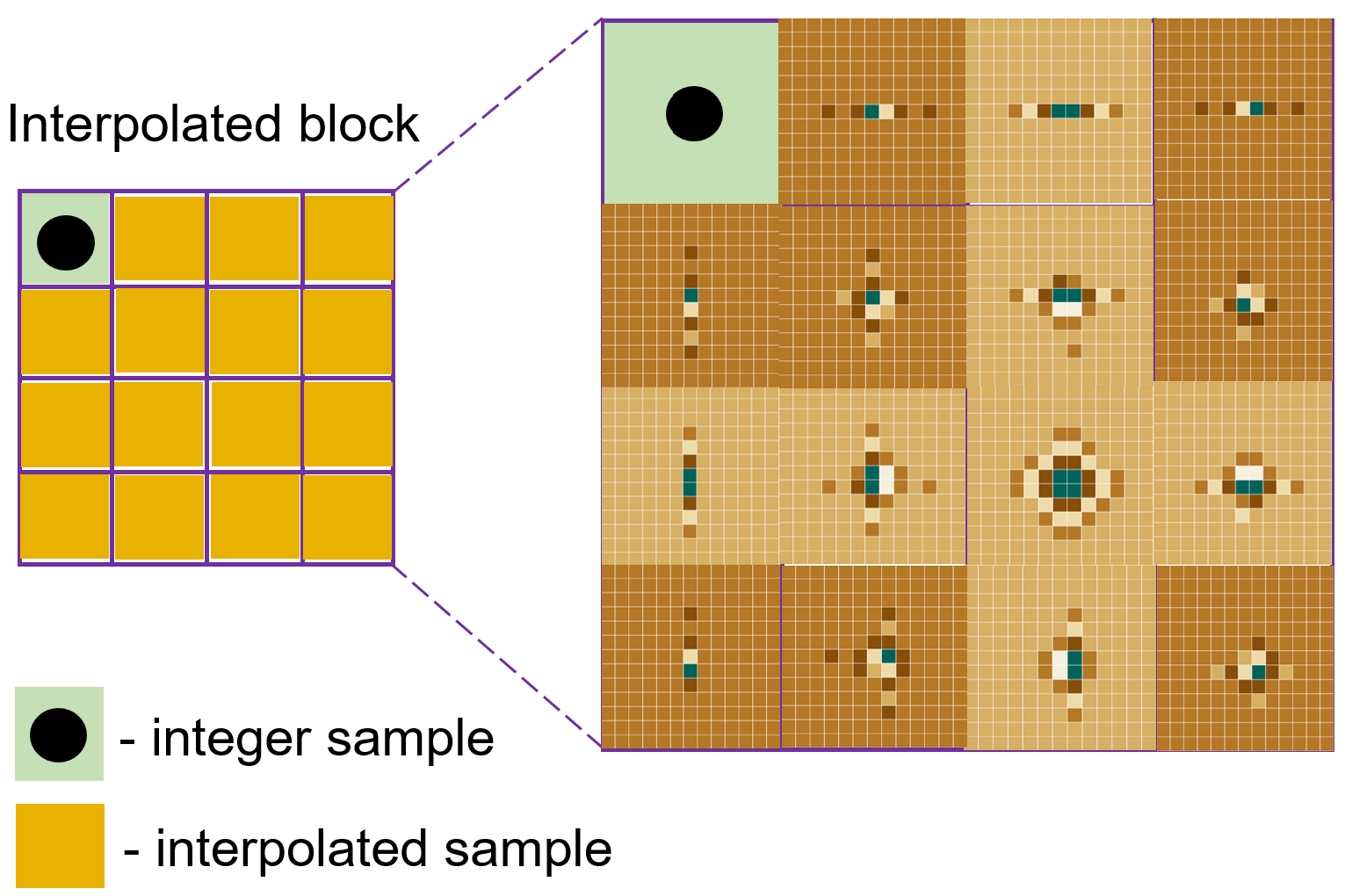}
    \caption{$15$ derived $13\times13$ interpolation filters, one for each quarter-pixel position }
    \label{fig:filters}
\end{figure}

Due to the network architecture of ScratchCNN, the described method directly computes samples of the resulting motion compensated prediction from the reference samples, instead of performing numerous convolutions defined by CNN layers. Furthermore, using this approach, it is possible to visually identify the contribution of each reference pixel in the $15$ interpreted filters, as illustrated in Fig. \ref{fig:filters}. 

\subsection{Encoding configuration}
\label{sec:encoding-configuration}
Tests for this preliminary work are done for simplified VVC inter-prediction, similar to HEVC conditions in \cite{Yan2019}. VVC Test Model (VTM) \cite{chen2018algorithmvtm3} version 6.0 was used as basis for this implementation. Common Test Conditions (CTCs) defined by JVET \cite{boyce2018ctc} were used, where these conditions were modified according to a number of restrictions imposed to encoder tools and algorithms. The flags include: Triangle=0, Affine=0, DMVR=0, BIO=0, WeightedPredP=0, WeightedPredB=0, MHIntra=0, SBT=0, MMVD=0, SMVD=0, IMV=0, SubPuMvp=0, TMVPMode=0; along with disabling the alternate half-pel interpolation filter and limiting VVC to quarter-pel fractional MC.

\subsection{Data generation and network training}
\label{sec:data-training}
The described modified VTM encoder was used to compress all frames in the \textit{BlowingBubbles} sequence, adopting the approach from \cite{Yan2019}. Additional restrictions were imposed to the encoding, to ensure that all blocks in the sequence are encoded using the same QP, and to ensure that only smaller coding units (CUs) are used. Training data was obtained using LDP configuration, with four different QPs of $22$, $27$, $32$ and $37$.

Although blocks equal or smaller than $16\times16$ samples were used for training in \cite{Yan2019}, VVC limits the minimum Coding Tree Unit (CTU) size to $32\times32$ luma samples, so blocks of maximum $32\times32$ samples were used when generating the data for training the evaluated approaches. Also different to HEVC, VVC uses a more complex partitioning scheme \cite{bross2019vvcdraft} which may result in non-square CUs. These rectangular blocks were also considered during the training of SRCNN and ScratchCNN.

Four sets (for QPs $22$, $27$, $32$ and $37$) of $15$ networks, one for each of the possible half-pel/quarter-pel positions in a $2$D space between $4$ integer pixels, were trained using Sum of Absolute Differences (SAD) as the loss function, along with the Adam optimiser. The approach is different from \cite{Yan2019}, where Mean Squared Error (MSE) and Stochastic Gradient Descent (SGD) were used as the loss function and optimiser.

\subsection{Integration into VVC}
After training the $60$ networks and extracting corresponding simplified $13\times13$ filter matrices from learned models, the filters were integrated within the VTM encoder as switchable interpolation filters. The selection between the conventional VVC filters and the $13\times13$ filters is performed at a CU level. One additional flag is correspondingly encoded in the bitstream and parsed by the decoder to determine which filter is used on a given block. Blocks coded in merge mode inherit usage of the same filter together with the merged motion information. The NN filters are only used for the luma component. If the QP of the CU is different to one of the $4$ QPs for which the filters are trained, then the filter trained for the closest QP to the current QP is used. Separate filters for bi-prediction were not considered at this stage.

\begin{table}
\footnotesize
\caption{Comparison of coding performances of different network structures for ClassD sequences, low-delay B (LDB) configuration, 32 frames.}
\label{tab:nn_comp}
\centering
\begin{tabular}{l|c|c|c} 
\hline
\hline
	& BD-Y [\%] & EncT [\%] & DecT [\%] \\
\hline
SRCNN & 0.67\% & 38915\% & 1322\% \\
\hline
\makecell[l]{ScratchCNN\\(MSE \& zero padding)} & 0.36\% & 859\% & 192\% \\
\hline
\makecell[l]{ScratchCNN\\(SAD \& no padding)} & -0.95\% & 863\% & 237\% \\
\hline
\hline
\end{tabular}
\end{table}

\section{Results}
\label{sec:results}

As mentioned in Section \ref{sec:encoding-configuration}, the proposed approach is tested for a modified VVC codec. Although neural networks are usually run on a GPU to enhance their run-time performance, all results reported here were obtained in a CPU environment.

Rather than integrating a deep learning software within VTM, all weights and biases ($8129$ parameters in total) are extracted from each of the $15$ trained SRCNNs and implemented in VTM as a series of matrix multiplications. In contrast, each trained ScratchCNN model is condensed in one $2$D matrix that contains $169$ parameters. As presented in Table \ref{tab:nn_comp} which summarises results for coding $32$ frames of ClassD sequences, the encoding time for a CPU implementation of SRCNN equals to $38915\%$ of the equivalent (restricted) VVC configuration. ScratchCNN encoding time increases are around $860\%$, showing a considerable running time reduction compared to SRCNN. Further comparisons are run for ScratchCNN trained in the way proposed in \cite{Yan2019} (MSE loss function, zero padding) against a network with SAD loss function, no padding on a block level, demonstrating how these changes bring significant coding gains.

\begin{table}
\footnotesize
\caption{Coding performance of the proposed approach for random access (RA), LDB and LDP configurations, entire sequence; BD-rate for luma.}
\label{tab:bdr_configs}
\centering
\begin{tabular}{cccccc} 

\hline
\hline
\multirow{2}{*}{\makecell{\textit{Sequence}\\\textbf{Class}}}                     & \multicolumn{5}{c}{Encoder configuration}\\
\cline{2-6}
& \multicolumn{1}{c}{RA} && \multicolumn{1}{c}{LDB} && \multicolumn{1}{c}{LDP}\\
\hline
\textit{BasketballDrill} \textbf{(C)} & -0.15\% && 0.11\% &&	-0.28\%\\								
\textit{BQMall} \textbf{(C)} & -0.32\% && -0.69\% &&	-1.25\%\\	
\textit{PartyScene} \textbf{(C)}	& -0.82\% && -1.92\% && -3.22\%\\
\textit{RaceHorses} \textbf{(C)} & 0.14\% && 0.19\% && 0.19\%\\	
\hline
\textbf{ClassC Overall} & \textbf{-0.29\%} && \textbf{-0.58\%} && \textbf{-1.14\%}\\
\hline
\textit{BasketballPass} \textbf{(D)} & -0.14\% && -0.33\% &&	-0.52\%\\	
\textit{BQSquare} \textbf{(D)} & -1.35\% && -3.02\% && -4.54\%\\
\textit{BlowingBubbles} \textbf{(D)} & -0.90\% && -2.18\% &&	-3.14\%\\	
\textit{RaceHorses} \textbf{(D)} & 0.04\% && 0.21\% && 0.02\%\\							
\hline
\textbf{ClassD Overall} & \textbf{-0.59\%} && \textbf{-1.33\%}	&& \textbf{-2.04\%}\\
\hline
\hline

\end{tabular}
\end{table}

\begin{table}
\footnotesize
\caption{Hit ratio for learned $13\times13$ interpolation filters, LDP configuration.}
\label{tab:hitratio}
\centering
\begin{tabular}{l|c|c|c|c} 
\hline
\hline
Class & QP 22 & QP 27 & QP 32 & QP 37 \\
\hline
Class C & 74.52\% & 85.28\% & 83.68\% & 80.62\%\\
\hline
Class D & 77.92\% & 88.35\% & 84.06\% & 79.66\% \\
\hline
\hline
\end{tabular}
\end{table}

Table \ref{tab:bdr_configs} summarises test results for a ScratchCNN switchable filter implementation within VTM 6.0 constrained conditions. As the network was trained on a Class D sequence, with its motion information extracted from an LDP configuration, the most significant coding gains are demonstrated for lower resolution test sequences. 
Since the learned $13\times13$ filters were implemented as switchable interpolation filters, each CU in VVC can select between the proposed NN $13\times13$ and conventional VVC filters during Rate-Distortion (R-D) optimisation. The ratio of CUs choosing the learned $13\times13$ filter across all CUs using sub-pixel MC is referred to as hit ratio. Hit ratio per QP of NN interpolation filters compared to VVC filters for both Class C and Class D sequences, LDP configuration is shown in Table \ref{tab:hitratio}, suggesting that the learned filters are performing well across all tested QPs.

The proposed approach achieves per class average and single configuration BD-rate saving of up to $2.0\%$ compared with the modified VVC, while significantly reducing the complexity of learned NN interpolation.

\section{Conclusions}
\label{sec:concl}
An approach for interpreting and understanding convolutional neural networks in visual data processing has been presented. The envisaged complexity reduction has been tested in the field of video coding, specifically on fractional-pixel motion compensation. Experimental results show a considerable encoder and decoder running time decrease when compared to previous state-of-the-art methods. Additional revisions to network training, such as using a SAD loss function and no padding, have also been proposed, displaying a notable increase in bitrate savings in a modified VVC encoder environment.

The presented work warrants further improvements, as Scratch-CNN's encoding time needs additional complexity reductions for possible future practical applications. Likewise, results need to be verified in VTM CTC. Greater diversity between the training and testing datasets is also required. Lastly, VVC uses a combination of SAD and a full R-D cost computation as a loss metric for motion estimation, meaning the neural network's SAD loss function currently doesn't describe the video coding loss metric in full.

\bibliographystyle{IEEEbib}
\bibliography{references}

\end{document}